\documentclass[11pt,prd,preprint,superscriptaddress]{revtex4}
\usepackage{revsymb}
\usepackage{graphicx}
\usepackage{bm}
\usepackage{amsmath}
\usepackage{amssymb}
\usepackage{graphicx}

\begin{document}

\title{Power-law solutions for TeVeS}

\author{C.E. Magalhaes Batista\footnote{E-mail: cedumagalhaes@hotmail.com}}
\affiliation{Universidade Federal do Esp\'{\i}rito Santo,
Departamento
de F\'{\i}sica\\
Av. Fernando Ferrari, 514, Campus de Goiabeiras, CEP 29075-910,
Vit\'oria, Esp\'{\i}rito Santo, Brazil}

\author{W. Zimdahl\footnote{E-mail: zimdahl@online.de}}
\affiliation{Universidade Federal do Esp\'{\i}rito Santo,
Departamento
de F\'{\i}sica\\
Av. Fernando Ferrari, 514, Campus de Goiabeiras, CEP 29075-910,
Vit\'oria, Esp\'{\i}rito Santo, Brazil}

\begin{abstract}
The dynamics of TeVeS in a homogeneous and isotropic universe is
shown to be equivalent to the dynamics of an interacting
two-component system, consisting of a scalar field and a
``fluid", related to the matter part, with explicitly given coupling term.  Scaling solutions (solutions with a constant ratio of the energy densities of both components) in the ``Einstein frame" are found which are exponentially expanding or contracting with no remaining freedom for Bekenstein's $F$ function. In the ``physical frame" these solutions are of the power-law type. An equivalent General Relativity (GR) picture of the dynamics suggests that it is the scalar field which plays the role of dark matter, while the ``matter" has to mimic (phantom-type) dark energy.
\end{abstract}


\maketitle
\date{\today}

\section{Introduction}
\label{Introduction}

The late-time cosmological acceleration, observationally suggested in \cite{Riess}, has been backed up over the years by growing direct and indirect evidence.
Direct evidence is provided by the luminosity-distance data of
supernovae of type Ia (SNIa) \cite{SNIa} (see, however, \cite{sarkar}), indirect evidence comes from the anisotropy spectrum of the cosmic microwave background radiation (CMBR) \cite{cmb}, from large-scale-structure data \cite{lss}, from the integrated Sachs--Wolfe effect
\cite{isw}, from baryonic acoustic
oscillations \cite{eisenstein} and from gravitational lensing \cite{weakl}.
An interpretation of these results within Einstein's GR requires the existence of an unknown dark sector that contributes roughly 95\% to the total cosmic energy density and manifests itself only through its gravitational action. Usually, this sector is divided into two components, dark matter (DM) and dark energy (DE), although there exist unified models of the cosmic substratum in which one single component shares propertied of both DM and DE (see \cite{unified} and references therein).
While the concept of DE is rather new, DM has been a topic both in cosmology and astrophysics for a long time. It was introduced by Zwicky as early as 1933 \cite{zwicky}. In a cosmological context DM is required to explain cosmic structure formation. It has to provide the potential wells for the baryonic perturbations after radiation decoupling. By itself, the fractional baryonic energy density perturbations would not have had sufficient time to enter the non-linear regime.
Despite of many attempts and the existence of several candidates the physical nature of DM remains a mystery until today (for recent reviews see \cite{einasto,bergstrom,primack}) . In many occasions its gravitational action in GR-based cosmology is simulated by a pressureless fluid component.
The failure of detecting DM over the decades has provoked more radical approaches which question the  universal validity of GR. The hope here is, that  within a (still to be found) modified theory of gravity DM may be superfluous. One remarkable step in this direction was Milgrom's MOND (Modified Newtonian Dynamics) \cite{milgrom}. The intention of MOND is to have visible matter as the only source of gravity and  to avoid the necessity of a dark matter component.
The main feature of this theory is that it modifies Newtonian gravity for accelerations below a threshold of the order of $10^{-10}\mathrm{m/s}^{2}$. MOND has been quite successful in describing galaxy rotation curves. For a recent review see \cite{milgromrev}. On the other hand, MOND is a non-relativistic theory and cannot  account for a cosmological dynamics. A relativistic version of MOND, called TeVeS (Tensor Vector Scalar), was developed by Bekenstein \cite{bekenstein}.

As are scalar-tensor theories, TeVeS is characterized by two different metrics. But while in scalar-tensor theories these metrics are related by a conformal transformation, the corresponding transformation in TeVeS is more complicated. In particular, it includes an additional unit time-like vector field. Moreover, there appears a free function in TeVeS, that is not determined by the theory. Einstein's GR can be recovered from TeVeS in a suitable limit. Cosmological applications of TeVeS were studied in \cite{hao,ratra,Skord,dodel,chiu,zhao,bourliot,zhao2,skordis,zunckel}. In several of these studies the question was raised whether TeVeS can also account for an accelerated expansion or, in other words, whether it can also avoid the explicit introduction of a dark energy component \cite{hao,zhao2}.
Reviews on TeVeS can be found, e.g., in \cite{skordisrev,ferrstark}.
A critical account of MOND-like gravity was given in \cite{bruneton}.

Because of its many degrees of freedom TeVeS is technically complicated. Postulating for the physical metric a structure of the Robertson-Walker type with spatially flat sections simplifies the formalism but the resulting dynamics is still not straightforwardly transparent.
It is the purpose of this paper to obtain simple analytic homogeneous and isotropic background solutions which allow for a discussion of at least some of the aspects of TeVeS. Our strategy here is similar to that already applied previously for scalar-tensor theories \cite{EdWi}. We shall make use of the circumstance that in the frame which is the analogue of the Einstein frame, the basic dynamics can be mapped onto the dynamics of an interacting two-component system. Then we look for scaling solutions of this system, i.e., for solutions with a constant ratio of the energy densities of both components.
For previous studies
of scaling solutions of the cosmological dynamics see, e.g.,  \cite{wetterich,uzan,ascale,amendola99,Wetterich,CoLiWands,Ferreira,ZBC}.
Subsequently, these solutions are transformed into the ``physical" frame, resulting in power-law solutions for the physical scale factor. Already known solutions are rediscovered as special cases in a different context.
Of course, power-law solution can only be considered as approximations for a specific epoch. In particular, they cannot describe any scenario, i.e., the transition between periods of different equations of state of the cosmic medium. Each of these phases has to be characterized by a different power-law solution, always under the condition that a power-law solutions is a reasonable approximation at all. Based on the explicitly known ``physical" frame dynamics of TeVeS we establish an effective GR equivalent and compare the emerging picture with the predictions of the $\Lambda$CDM model. The perhaps surprising results of our calculations are: (i) an early matter phase can only be recovered
in the presence of an additional phantom-type dark energy, although the latter is dynamically negligible.
(ii) To be consistent with the observations, a separate phantom dark-energy component is required for the  present Universe. In our solution  TeVeS does not geometrize dark energy.
But in accordance with the general intention of MOND type theories, no separate dark matter is necessary. TeVeS can geometrize cosmological dark matter, but only in the presence of an additional phantom dark energy component.
While it remains open whether or not solutions of this type are (at least approximately) realized in our Universe, we hope that they may at least shed some light on the structure of TeVeS.

\noindent The paper is organized as follows. Section \ref{General}
recalls the basic structure of TeVeS. In section \ref{Basics} the
general theory is applied to homogeneous and isotropic cosmology.
In section \ref{two-component} we map the corresponding dynamics
on the dynamics of an interacting two-component system with a
coupling term that is explicitly prescribed by the structure of
TeVeS. In section \ref{scaling} we find scaling solutions for this
system, i.e. solutions with a constant ratio of the energy
densities of both components. An effective GR description of the cosmic medium is given in \ref{cosmed}.  Section \ref{discussion} summarizes
the results of the paper.

\section{General theory}
\label{General}

The dynamical quantities in Bekenstein's TeVeS  are Einstein's
metric tensor $g_{\mu\nu}$, a timelike unity four-vector
$U_{\alpha}$ and a scalar field $\phi$. The physical metric tensor
$\tilde{g}_{\mu\nu}$ is constructed out of these quantities by
\begin{equation}
\tilde{g}_{\mu\nu} = e^{-2\phi}g_{\mu\nu} -
2U_{\mu}U_{\nu}\sinh\left(2\phi\right) =
e^{-2\phi}\left(g_{\mu\nu} + U_{\mu}U_{\nu}\right) - e^{2\phi}
U_{\mu}U_{\nu}
 \ .\label{tilg}
\end{equation}
The action of the system is
\begin{equation}
S = S_{g} + S_{s} + S_{v} + S_{m}
 \ ,\label{S}
\end{equation}
where
\begin{equation}
S_{g} = \frac{1}{16\pi G}\int d^{4}x \sqrt{-g}g^{\mu\nu}R_{\mu\nu}
 \ ,\label{Sg}
\end{equation}
with Newton's gravitational constant $G$,
\begin{equation}
S_{s} = -\frac{1}{2}\int d^{4}x
\sqrt{-g}\left[\sigma^{2}h^{\mu\nu}\phi_{,\mu}\phi_{,\nu} +
\frac{G}{2l^{2}}\sigma^{4}F(kG\sigma^{2})\right]
 \ ,\label{Ss}
\end{equation}
where $\sigma$ is a non-dynamical scalar field, $k$ is a
dimensionless constant, $l$ is a constant length scale and
$h^{\mu\nu} = g^{\mu\nu} - U^{\mu}U^{\nu}$ (the use of $h^{\mu\nu}$ instead of $g^{\mu\nu}$ avoids superluminal propagation of perturbations) ,
\begin{equation}
S_{v} = -\frac{K}{32}\pi G\int d^{4}x
\sqrt{-g}\left[g^{\alpha\beta}g^{\mu\nu}U_{\left[\alpha,\mu\right]}U_{\left[\beta,\nu\right]}
- \frac{2\lambda}{K}\left(g^{\mu\nu}U_{\mu}U_{\nu} +
1\right)\right]
 \ ,\label{Sv}
\end{equation}
where $K$ is another dimensionless constant and $\lambda$ is a
Lagrangean multiplier. The square brackets in
$U_{\left[\alpha,\mu\right]}$ denote antisymmetrization. Finally,
$S_{m}$ is the matter part of the action,
\begin{equation}
S_{m} = \int d^{4}x \sqrt{-\tilde{g}}L_{m}
 \ ,\label{Sm}
\end{equation}
with the matter Lagrangean $L_{m} = L_{m}\left(\tilde{g}_{\mu\nu},
...\right)$. ``Matter" is understood here in a general sense. In particular, it can also be
a dark energy component.
The basic dynamic equations of TeVeS are obtained by
varying the action (\ref{S}) with respect to $U_{\mu}$, $\sigma$,
$\phi$, $g^{\mu\nu}$ and $\lambda$. In general form they are
given, e.g., in \cite{bekenstein}. Here we are interested in the
spatially homogeneous and isotropic case for which the basic set
of equations considerably simplifies.

\section{Basics of homogeneous and isotropic cosmology}
\label{Basics}

The physical metric is assumed to be of the spatially flat FLRW type,
\begin{equation}
d\tilde{s}^{2} = - d\tilde{t}^{2} + \tilde{a}^{2}\left[dr^{2} +
r^{2}\left(d\theta^{2} + \sin^{2}\theta d
\varphi^{2}\right)\right]
 \ .\label{dts}
\end{equation}
The conversion relations between the quantities with tilde (physical frame) and without
tilde (Einstein frame) are
\begin{equation}
d\tilde{t} = \exp\left[\phi\right]d t\ ,\qquad \tilde{a} = \exp
\left[-\phi\right] a
 \ .\label{conv}
\end{equation}
With the definitions
\begin{equation}
H \equiv \frac{\frac{da}{dt}}{a}\ ,\qquad \tilde{H} \equiv \frac{\frac{d
\tilde{a}}{d \tilde{t}}}{\tilde{a}}
 \ ,\label{defH}
\end{equation}
the relation between $H$, the Hubble rate in the Einstein frame and $\tilde{H}$, the Hubble rate in the physical frame, is
\begin{equation}
\tilde{H} = \exp[-\phi]\left(H - \frac{d\phi}{dt}\right)
 \ .\label{tH}
\end{equation}
The matter part is assumed to be describable  by the
energy-momentum tensor of a perfect fluid,
\begin{equation}
\tilde{T}_{\alpha\beta} =
\tilde{\rho}\tilde{u}_{\alpha}\tilde{u}_{\beta} +
\tilde{p}\left(\tilde{g}_{\alpha\beta} +
\tilde{u}_{\alpha}\tilde{u}_{\beta}\right)
 \ .\label{tilT}
\end{equation}
At this stage it is left open, which type of matter $\tilde{T}_{\alpha\beta}$ is supposed to describe. As already mentioned, it may also have a negative equation-of-state (EoS) parameter and account for some form of DE.
The matter four-velocity $\tilde{u}_{\mu}$ and the vector $U_{\mu}$ are related by $\tilde{u}_{\mu} =
e^{\phi} U_{\mu}$ with $U^{\mu} = \delta^{\mu}_{t}$
\cite{bekenstein}. The matter energy density is
\begin{equation}
\rho_{m} \equiv \tilde{T}_{00} = \tilde{\rho}\,\exp[2\phi]
 \ ,\label{rhom}
\end{equation}
the matter pressure
\begin{equation}
p_{m} =  \tilde{p}\,\exp[-2\phi]
 \ .\label{pm?}
\end{equation}

Varying the action (\ref{S}) with respect to $g_{\mu\nu}$ results,
when specified to the spatially homogeneous and isotropic case
(\ref{dts}), in a Friedmann type equation for the Einstein frame scale factor
$a$,
\begin{equation}
\frac{1}{a^{2}}\left(\frac{da}{dt}\right)^{2} = \frac{8 \pi G}{3}\,\rho_{eff}
 \ \label{Friedm}
\end{equation}
with an effective energy density
\begin{equation}
\rho_{eff} = \tilde{\rho}\,\exp[-2\phi]  +
\sigma^{2}\left(\frac{d\phi}{dt}\right)^{2} +
\frac{G\sigma^{4}}{4l^{2}}\,F(kG\sigma^{2})
 \ \label{rhoeffgr}
\end{equation}
and in
\begin{equation}
\frac{2}{a}\frac{d^{2}a}{dt^{2}} + \frac{1}{a^{2}}\left(\frac{da}{dt}\right)^{2} = - 8 \pi
G\,p_{eff}
 \ ,\label{dda}
\end{equation}
with an effective pressure
\begin{equation}
p_{eff} =  \tilde{p}\,\exp[-2\phi] + \sigma^{2}\left(\frac{d\phi}{dt}\right)^{2} -
\frac{G\sigma^{4}}{4l^{2}}\,F(kG\sigma^{2})
 \ .\label{peffgr}
\end{equation}
The physical matter energy density $\tilde{\rho}$ obeys the equation
\begin{equation}
\frac{d\tilde{\rho}}{dt} = 3\left(\frac{d\phi}{dt} -
\frac{1}{a}\frac{da}{dt}\right)\left(\tilde{\rho} + \tilde{p}\right)
 \ ,\label{dottrho}
\end{equation}
which is the rewritten conservation equation
\begin{equation}
\frac{d \tilde{\rho}}{d \tilde{t}} = - 3
\tilde{H}\left(\tilde{\rho} + \tilde{p}\right)
 \ \label{tildecons}
\end{equation}
of the physical frame.
The scalar field equation, obtained by the variation of (\ref{S})
with respect to $\phi$, specifies to
\begin{equation}
\frac{d}{dt}\left(\sigma^{2}\frac{d\phi}{dt}\right) + 3 H
\sigma^{2} \frac{d\phi}{dt} =  - \frac{1}{2}\,\exp[- 2\phi]
\left(\tilde{\rho} + 3 \tilde{p}\right)
 \ . \label{ddphir2}
\end{equation}
Finally, varying (\ref{S}) with respect to $\sigma$ yields
\begin{equation}
\frac{G}{4 l^{2}} \frac{d}{d\sigma^{2}}\left(\sigma^{4} F\right) =
\left(\frac{d\phi}{dt}\right)^{2}
 \  .\label{consist}
\end{equation}
Equations (\ref{Friedm}), (\ref{dda}), (\ref{ddphir2}) and
(\ref{consist}) are the basic equations of TeVeS in a homogeneous
and isotropic, spatially flat universe.

\section{A two-component description}
\label{two-component}

The structures of the expressions (\ref{rhoeffgr}) and (\ref{peffgr}) suggest a
split (cf. \cite{hao})
\begin{equation}
\rho_{eff} = \rho_{\phi} + \rho_{f}
 \ \label{rhoeffdec}
\end{equation}
and
\begin{equation}
p_{eff} =  p_{\phi} + p_{f}
 \ ,\label{peffdec}
\end{equation}
respectively, with the field part (subscript $\phi$)
\begin{equation}
\rho_{\phi} = \sigma^{2}\left(\frac{d\phi}{dt}\right)^{2} + \frac{G\sigma^{4}}{4
l^{2}} \,F
 \ \label{rho1}
\end{equation}
and a fluid part (subscript $f$)
\begin{equation}
\rho_{f} = \tilde{\rho}\,\exp[-2\phi] =\rho_{m}\,\exp[-4\phi]
 \ ,\label{rho2}
\end{equation}
as well as
\begin{equation}
p_{\phi} = \sigma^{2}\left(\frac{d\phi}{dt}\right)^{2} - \frac{G\sigma^{4}}{4 l^{2}}
\,F
 \ \label{p1}
\end{equation}
and
\begin{equation}
p_{f} =  \tilde{p}\,\exp[-2\phi] = p_{m}
 \ .\label{p20}
\end{equation}
The fluid EoS parameter is
\begin{equation}
\frac{p_{f}}{\rho_{f}}=
\frac{\tilde{p}\,\exp[-2\phi]}{\tilde{\rho}\,\exp[-2\phi])} =
\frac{\tilde{p}}{\tilde{\rho}} \equiv \tilde{w}
 \ .\label{p2}
\end{equation}
We shall restrict ourselves here to constant values of the fluid
EoS parameter $\tilde{w}$. The field EoS
parameter is
\begin{equation}
w_{\phi} \equiv \frac{p_{\phi}}{\rho_{\phi}} =
\frac{\sigma^{2}\left(\frac{d\phi}{dt}\right)^{2} - \frac{G}{4
l^{2}}\sigma^{4}F}{\sigma^{2}\left(\frac{d\phi}{dt}\right)^{2} + \frac{G}{4
l^{2}}\sigma^{4}F}
 \ .\label{w1}
\end{equation}
We emphasize (cf. \cite{ratra}), that $\rho_{f}$ does not coincide
with the matter energy density which is $\rho_{m} =
\tilde{T}_{00}$ (cf. Eq.~(\ref{rhom})). Consequently, the
decomposition (\ref{rhoeffdec}) is not a split into field and
matter parts. The theory does not admit a clear separation between
the both. Also the fluid parts (\ref{rho2}) and (\ref{p2}) do
depend on $\phi$.
However, according to (\ref{p2}), the ratio $\frac{p_{f}}{\rho_{f}}$ coincides with the physical EoS parameter $\tilde{w} = \frac{\tilde{p}}{\tilde{\rho}}$. As already mentioned, negative values of $\tilde{w}$ are admitted here.

The coupling between the components can be made
transparent by considering the corresponding energy balances. With
the help of (\ref{consist}) the dynamics of the previous section
is equivalent to
\begin{equation}
\frac{d\rho_{\phi}}{dt} + \frac{3}{a}\frac{da}{dt}\left(1 +
w_{\phi}\right)\rho_{\phi} = - \frac{d\phi}{dt}\,\left(1 +
3\tilde{w}\right)\rho_{f}
 \ \label{dotrho1}
\end{equation}
and
\begin{equation}
\frac{d\rho_{f}}{dt} + \frac{3}{a}\frac{da}{dt}\left(1 + \tilde{w}\right)\rho_{f}
= \frac{d\phi}{dt}\,\left(1 + 3\tilde{w}\right)\rho_{f}
 \ . \label{dotrho2}
\end{equation}
The individual balance equations (\ref{dotrho1}) and
(\ref{dotrho2}) add up to the total energy conservation
\begin{equation}
\frac{d\rho_{eff}}{dt} + \frac{3}{a}\frac{da}{dt}\left(\rho_{eff} + p
_{eff}\right) = 0 \  .\label{dotrhoeff}
\end{equation}
It is the system of equations (\ref{dotrho1}) and (\ref{dotrho2})
which we are going to study in the following. This system consists
of two components which are interacting with each other. The
interaction is given explicitly by the right-hand sides of
equations (\ref{dotrho1}) and (\ref{dotrho2}).
It vanishes in the particular case $\tilde{w} = -\frac{1}{3}$.
It is expedient to
emphasize that equation (\ref{consist}) (sometimes neglected in
the literature \cite{ratra}) was used in order to arrive at the
system (\ref{dotrho1}) and (\ref{dotrho2}). We shall solve the
system of equations (\ref{dotrho1}) and (\ref{dotrho2}) under the
condition  $\tilde{w} = $ constant. Subsequently, we shall also assume
$w_{\phi} = $ constant as well. Our aim then will be to look for scaling solutions,
i.e. solutions, for which the energy density ratio
$\frac{\rho_{f}}{\rho_{\phi}}$ is constant.

\section{A scaling solution}
\label{scaling}

\subsection{General properties}

For $\tilde{w} = $ constant, equation (\ref{dotrho2}) can be solved
immediately,
\begin{equation}
\rho_{f} = \rho_{f0} \,\exp[\left(1 + 3\tilde{w}\right)\phi
]\,\left(\frac{a_{0}}{a}\right)^{3\left(1 + \tilde{w}\right)}
 \ , \label{rho2sol}
\end{equation}
where the subscript $0$ denotes some reference value. Via
(\ref{rho2}) and (\ref{rhom}) corresponding expressions for
$\rho_{m}$ and $\tilde{\rho}$, respectively, are straightforwardly
obtained. The latter reproduces the solution (24) in \cite{ratra}.
Defining a ratio $R$ by
\begin{equation}
R \equiv \frac{\rho_{f}}{\rho_{\phi}}
 \  , \label{R}
\end{equation}
the balance (\ref{dotrho1}) becomes
\begin{equation}
\frac{1}{\rho_{\phi}}\frac{d\rho_{\phi}}{dt} + \frac{3}{a}\frac{da}{dt}\left(1
+ w_{\phi}\right) = - \frac{d\phi}{dt}\,\left(1 + 3\tilde{w}\right)R
 \ . \label{dotrho1R}
\end{equation}
For the simplest case that both $w_{\phi}$ and $R$ are constants,
equation (\ref{dotrho1R}) has the solution
\begin{equation}
\rho_{\phi} = \rho_{\phi 0} e^{-\left(1 + 3\tilde{w}\right)\phi
R}\,\left(\frac{a_{0}}{a}\right)^{3\left(1 + w_{\phi}\right)}
 \ . \label{rho1Rsol}
\end{equation}
From now on we shall restrict our discussion to this case, i.e.,
to constant values of $w_{\phi}$ and $R$.
While constant EoS parameters can certainly not account for a continuous transition between different epochs of the cosmological evolution such as the transition from decelerated to accelerated expansion, they are useful as simple exact solutions which are valid piecewise.
Scaling solutions of the cosmological dynamics, i.e., solutions with a constant ratio of the energy densities of the dynamically relevant components, have been studied, e.g., in \cite{ascale,wetterich,uzan,amendola99,Wetterich,CoLiWands,Ferreira,ZBC}.
We mention also that the ratio of the energy density of the scalar field to that of the matter, corresponding to $R^{-1}$ in our notation, turned out to be approximately constant for Bekenstein's choice of the free function $F$ (cf. \cite{skordisrev}). This may be seen as an additional motivation to investigate such kind of solutions. It will turn out, however, that the requirement of a constant $R$ does not reproduce Bekenstein's choice of $F$ but leads to a different expression. Studying scaling solutions of the mentioned type also implies that we neglect the baryon component of the Universe since its contribution to the total energy budget is known to be small and the baryons as well as the also omitted photons are not expected to participate in the scaling of the dynamically dominating components.

Combination of (\ref{rho2sol}) and (\ref{rho1Rsol}) yields
\begin{equation}
R  = \frac{\rho_{f 0}}{\rho_{\phi 0}}\,\,e^{\left(1 +
3\tilde{w}\right)\phi\left(1 +
R\right)}\left(\frac{a_{0}}{a}\right)^{3\left(\tilde{w} -
w_{\phi}\right)}
 \ . \label{R=}
\end{equation}
For this expression to be constant we may consider two cases.

\noindent (i) The ratio $R$ in (\ref{R=}) is constant for $\phi = $ const and
$\tilde{w} = w_{\phi}$. Since $\phi = $ const implies $w_{\phi} = -1$
(cf.~(\ref{w1})), we have also $\tilde{w} = -1$. Then $\rho_{\phi}$,
$\rho_{f}$ and $\rho_{eff}$ are constant and $a \propto e\,^{\pm H
t}$ with $H = $ const.
Now we take  into account the scalar field equation
(\ref{ddphir2}) which can be written as
\begin{equation}
\frac{d}{dt}\left(\sigma^{2}\frac{d\phi}{dt}\right) + 3 H
\sigma^{2} \frac{d\phi}{dt} =  - \frac{1}{2}\,\left(1 +
3\tilde{w}\right)\rho_{f}
 \ . \label{ddphir2+}
\end{equation}
An exact solution for $\phi = $ const requires $\rho_{f} = 0 \
\rightarrow \ R=0$ (unless $\tilde{w} = - \frac{1}{3}$), which means it is a vacuum solution. This
solution reproduces the relations (22) and (23) in \cite{ratra}. However, there
exists another solution for $\rho_{f} = 0$, namely $\sigma^{2} =
0$. As we shall see, this solution corresponds to $w_{\phi} = -1$.
The point here is, that in the limit $R \rightarrow 0$ it follows
that $\sigma^{2} \rightarrow  0$, but, on the other hand,
$\sigma^{4}F \rightarrow  $ const as will be shown below. Under
these conditions (\ref{w1}) leads to $w_{\phi} = -1$.

\noindent (ii) The second case for which $R$ in (\ref{R=}) is
constant is (we assume  $a_{0} = 1$ from now on)
\begin{equation}
e^{\left(1 + 3\tilde{w}\right)\phi\left(1 + R\right)} =
a^{3\left(\tilde{w} - w_{\phi}\right)} \quad\Rightarrow \quad
e^{\left(1 + 3\tilde{w}\right)\phi} = a^{3 \frac{\tilde{w} - w_{\phi}}{1 +
R}}
 \ . \label{R=2}
\end{equation}
This relation fixes the dynamics of $\phi$. It implies
\begin{equation}
\frac{d\phi}{dt} = 3 \frac{\tilde{w} - w_{\phi}}{\left(1 +
3\tilde{w}\right)\left(1 + R\right)}\,H
 \ , \label{dotphiH}
\end{equation}
i.e., a proportionality between $\frac{d\phi}{dt}$ and $H$. Again we exclude here the case $\tilde{w}= -\frac{1}{3}$. Using
(\ref{R=2}) in the expressions (\ref{rho1Rsol}) and
(\ref{rho2sol}) for $\rho_{\phi}$ and $\rho_{f}$, respectively, we
find the scaling solution
\begin{equation}
\rho_{\phi} \propto \rho_{f} \propto \rho_{eff} \propto a^{- 3
\frac{1 + w_{\phi} + R\left(1 + \tilde{w}\right)}{1 + R}}
 \ . \label{rhopropto1}
\end{equation}
Via Friedmann's equation (\ref{Friedm}) it follows that
\begin{equation}
H \propto
\pm\, a^{- \frac{3}{2}\frac{1 + w_{\phi} + R\left(1 +
\tilde{w}\right)}{1 + R}}
 \ . \label{rhopropto}
\end{equation}
Notice that both signs are possible here. There are both expanding
$(H>0)$ and contracting $(H<0)$ solutions.
Integrating (\ref{rhopropto}), we obtain
\begin{equation}
a \propto t^{\frac{2}{3\left(1 + w_{eff}\right)}}\ , \qquad w_{eff} \equiv \frac{w_{\phi} + R \tilde{w}}{1+R}
 \ . \label{at}
\end{equation}
$w_{eff}$ is the EoS parameter in the Einstein frame. Notice that it has the same structure as a corresponding term in scalar-tensor theories \cite{EdWi}.
From (\ref{R=2}) we have ($\tilde{w} \neq - \frac{1}{3}$)
\begin{equation}
e^{\phi} = a^{3 \frac{\tilde{w}-w_{eff}}{1+3\tilde{w}}} \quad \Rightarrow\quad
e^{\phi} \propto t^{\ 2 \frac{\tilde{w}-w_{eff}}{\left(1+w_{eff}\right)\left(1+3\tilde{w}\right)}}
 \ . \label{ephia}
\end{equation}
Further constraints can again be obtained from the scalar field
equation (\ref{ddphir2}), which is now written as
\begin{equation}
\frac{d}{dt}\left(\sigma^{2}\frac{d\phi}{dt}a^{3}\right) = -
\frac{1}{2}\,\left(1 + 3\tilde{w}\right)\, a^{3}\,\rho_{f}
 \ . \label{ddphia}
\end{equation}
Changing on the left-hand side the variables from $t$ to $a$, we
obtain
\begin{equation}
\frac{d}{d a} \left(\sigma^{2}\frac{d\phi}{dt} a^{3}\right) =  -
\frac{1}{2}\,\left(1 + 3\tilde{w}\right)\, \frac{a^{2}}{H}\,\rho_{f}
 \ . \label{ddphiH}
\end{equation}
With $H$ from (\ref{rhopropto}) and $\rho_{f}$ from (\ref{rhopropto1}),
integration of (\ref{ddphiH}) yields
\begin{equation}
\sigma^{2}\frac{d\phi}{dt} =  - \frac{1}{2}\,\left(1 + 3\tilde{w}\right)\,
\frac{\rho_{f 0}}{H_{0}}\,\frac{1 + R}{\frac{3}{2}\left[1 -
w_{\phi} + R\left(1 - \tilde{w}\right)\right]}a^{- \frac{3}{2} \frac{1
+ w_{\phi} + R\left(1 + \tilde{w}\right)}{1 + R}}
 \ ,  \label{mudotphi}
\end{equation}
where $\rho_{f 0}$ and $H_{0}$ are the values for $\rho_{f}$ and $H$ respectively, for $a=1$.
Taking into account also
(\ref{rhopropto}) and $3 H_{0}^{2} = 8 \pi G\rho_{eff 0} $, relation
(\ref{mudotphi}) can be transformed into
\begin{equation}
8 \pi G \sigma^{2} \frac{d\phi}{dt} =  - \frac{\left(1 +
3\tilde{w}\right)R}{1 - w_{\phi} + R\left(1 - \tilde{w}\right)}\,\,H
 \ . \label{dotphikH}
\end{equation}
Comparing the previously found relation (\ref{dotphiH}) between
$\frac{d\phi}{dt}$ and $H$ with (\ref{dotphikH}), allows us to obtain
\begin{equation}
\frac{8 \pi G}{3}\sigma^{2} = -
\frac{R}{9}\frac{\left(1+R\right)\left(1 +
3\tilde{w}\right)^{2}}{\left(\tilde{w} - w_{\phi}\right)\left[1 - w_{\phi}
+ R\left(1 - \tilde{w}\right)\right]}
 \ . \label{mu/k}
\end{equation}
Equation (\ref{mu/k}) implies that the parameter $\sigma^{2}$ is
constant. Now, if $\sigma^{2}$ is constant, the function $F$ is
constant as well. Via (\ref{consist}), a constant $F$ means that
$\frac{d\phi}{dt} = $ const. But with $F = $ constant and $\frac{d\phi}{dt} =
$ constant, the  energy density (\ref{rho1}) is constant, hence
(\ref{rho2}) and (\ref{rhoeffdec}) are constant.
 This property leads to a condition on the solution
(\ref{rhopropto1}) namely,
\begin{equation}
\rho_{eff} \propto a^{- 3 \frac{1 + w_{\phi} + R\left(1 +
\tilde{w}\right)}{1 + R}} = \mathrm{const} \quad\Rightarrow \quad
w_{\phi} = - 1 - R\left(1 + \tilde{w}\right)
 \ . \label{rhoconst}
\end{equation}
Inserting (\ref{rhoconst}) into (\ref{mu/k}) the latter becomes
\begin{equation}
\frac{8 \pi G}{3}\sigma^{2} = - \frac{R}{18}\frac{\left(1 +
3\tilde{w}\right)^{2}}{\left(1 + \tilde{w}\right)\left(1 + R\right)}
 \ . \label{mu/kfin}
\end{equation}
The quantity $\sigma^{2}$ is related to the fraction and the
EoS of the matter. It diverges for $\tilde{w} = -1$. In the vacuum limit $R =
0$ we recover the previously mentioned solution $\sigma^{2} = 0$
(see the discussion following eq.~(\ref{ddphir2+})).

Now we consider the case $\tilde{w} = - \frac{1}{3}$ for which the equations (\ref{dotrho1}) and (\ref{dotrho2}) decouple. The solutions for $\rho_{f}$ and $\rho_{\phi}$ then are $\rho_{f} \propto a^{-2}$ and $\rho_{\phi} \propto a^{-3\left(1 + w_{\phi}\right)}$, respectively. In such a case a constant energy density ratio requires $w_{\phi} =- \frac{1}{3}$ as well. But with $w_{\phi} = \tilde{w} = - \frac{1}{3}$
the last relation in (\ref{rhoconst}) cannot be satisfied for a non-negative ratio $R$. Consequently, there does not exist a scaling solution of the type discussed here for $\tilde{w} = - \frac{1}{3}$. This case will be excluded from now on.

Use of the last relation of (\ref{rhoconst}) in the expression
(\ref{at}) for $w_{eff}$, consistently yields $w_{eff}= - 1$ together with $H= $ constant (cf.Eq.~(\ref{rhopropto})). Then $a\propto \exp\left[Ht\right]$, replacing (\ref{at}), and the relations (\ref{ephia})
reduce to
\begin{equation}
e^{\phi} = a^{3 \frac{1+\tilde{w}}{1+3\tilde{w}}} \propto e^{3H\frac{1+\tilde{w}}{1+3\tilde{w}}t}  \ .\label{ephi}
\end{equation}
If we use (\ref{rhoconst}) in (\ref{dotphiH}) we obtain
\begin{equation}
\frac{d\phi}{dt} = 3 \frac{1 + \tilde{w}}{1 + 3\tilde{w}}\,H
 \ . \label{dotphiHfin}
\end{equation}
The first relation (\ref{ephi}) reproduces equation (25) in
\cite{ratra}, which was introduced there as an ansatz to simplify
the dynamics without a further reasoning. Here it appears as a property of our scaling solution.
For $\tilde{w} > -\frac{1}{3}$ as well as for $\tilde{w} < -1$ the quantities $\frac{d\phi}{dt}$ and $H$ have the same sign, for $ - 1 < \tilde{w} < -\frac{1}{3}$ the sign is different.
Relation (\ref{dotphiHfin}) is valid for any constant value of $R$.

From
(\ref{mu/kfin}) we can draw the conclusion that $\sigma^{2}$ is
only positive for $\tilde{w}<-1$. This is exactly the condition under
which the authors of \cite{ratra} obtained exponential expansion.
A negative  $\sigma^{2}$, necessarily obtained for $\tilde{w} > -1$,
makes the kinetic term in (\ref{rho1}) negative, thus describing
phantom energy. This is consistent with (\ref{rhoconst}), where
$w_{\phi} < - 1$ implies $\tilde{w} > - 1$.
In particular, this property holds also also for a non-relativistic matter EoS $\tilde{w} = 0$.
As we shall see shortly,
despite of the negative kinetic term,  the energy densities of the
components and the total energy density are always positive.

Phantom type solutions (within GR) are known to suffer from
stability problems \cite{carroll,hsu}. On the other hand, there
are suggestions from higher dimensional theories that a phantom
character ``on the brane" may be an effective phenomenon
\cite{sahni,lue}. To the best of our our knowledge phantom
solutions for TeVeS and their properties have not been discussed
so far.
%Given that our scaling solution includes elements of a
%realistic description of the Universe, it would be necessary to
%show, how it is dynamically approached. Very likely, this would
%imply also a transition from $\sigma^{2}>0$ at high redshifts to
%$\sigma^{2}<0$ of our scaling solution.
In any case, since a
phantom component is not excluded observationally, this issue
deserves attention.

Another interesting special case is $w_{\phi} = 0$, equivalent to $\tilde{w} < - 1$.  Under this condition both components change their roles compared with the case $\tilde{w} = 0$. Now, the scalar field mimics matter (although not separately conserved), while the fluid component is of the phantom type.
Since TeVeS is a relativistic MOND-type theory and the intention of MOND was to make dark matter superfluous, it is now the scalar field itself which replaces DM. The ``fluid" on the other hand, has to be identified with a DE component. We emphasize, however, that all these considerations refer to the dynamics in the Einstein frame.

\subsection{The equation of state}

\noindent For the total EoS we find, using
(\ref{rhoconst}),
\begin{equation}
\frac{p_{eff}}{\rho_{eff}} = \frac{p_{\phi} + p_{f}}{\rho_{\phi} +
\rho_{f}} = \frac{1}{1 + R}w_{\phi} + \frac{R}{1
+ R}\tilde{w} = -1
 \  .\label{wtot2}
\end{equation}
The total EoS parameter is always $-1$, consistent
with $\rho_{eff} = $ const. Assuming the matter to behave non-relativistically, i.e.,
$\tilde{w} = 0$, corresponding to $w_{\phi} = -1 - R$ (cf.~(\ref{w1})),
we find from (\ref{mu/kfin}) and (\ref{dotphiHfin})
that
\begin{equation}
\frac{8 \pi G}{3}\sigma^{2} = - \frac{1}{18}\frac{R}{1 + R}  \ \quad \mathrm{and} \quad\
\frac{d\phi}{dt} = 3 \,H
\qquad\qquad (\tilde{w}=0)
 \ . \label{mu/kfin0}
\end{equation}
The corresponding relations for the special case $w_{\phi} = 0$, equivalent to $\tilde{w} = -1 - 1/R$, are
\begin{equation}
\frac{8 \pi G}{3}\sigma^{2} =  \frac{1}{2}\frac{\left(1 + \frac{2}{3}R\right)^{2}}{1 + R}
 \ \quad \mathrm{and} \quad\ \frac{d\phi}{dt} = \frac{1}{1 + 2R/3} \,H
\qquad\qquad (w_{\phi}=0)\ .
 \ \label{mu/kfin0+}
\end{equation}
With $R=0$ one realizes the existence of a vacuum solution $\rho_{\phi} =$ const of Eq.~(\ref{dotrho1}) also for $w_{\phi}=0$.

Friedmann's equation (\ref{Friedm}) can be written
$
3 H^{2} = 8\pi G\left[\rho_{\phi} + \rho_{f}\right]$
with $\rho_{\phi}$ and $\rho_{f}$ given by (\ref{rho1}) and
(\ref{rho2}), respectively. Then, with (\ref{R}), it is equivalent to
\begin{equation}
3 H^{2} = 8\pi G\,\left(1 +
R\right)\,\left[\sigma^{2}\left(\frac{d\phi}{dt}\right)^{2} +
\frac{G}{4l^{2}}\,\sigma^{4}\,F\right]
 \ . \label{FriedmR}
\end{equation}
Upon combining (\ref{mu/kfin}) and (\ref{dotphiHfin}) we arrive at
\begin{equation}
\sigma^{2}\left(\frac{d\phi}{dt}\right)^{2}  = -\frac{3}{8\pi G}\frac{R}{2}\frac{1 +
\tilde{w}}{1 + R}H^{2}
 \ \label{kin}
\end{equation}
for the first term in the bracket in (\ref{FriedmR}).
 Introducing
(\ref{kin}) in (\ref{FriedmR}) leads to
\begin{equation}
8 \pi G\frac{G}{4l^{2}}\,\sigma^{4}\,F = 3H^{2}\, \frac{1 +
\frac{R}{2}\left(1 + \tilde{w}\right)}{1 + R}\,
 \ . \label{FriedmR1}
\end{equation}
Combination of (\ref{kin}) and (\ref{FriedmR1}) results in
\begin{equation}
\sigma^{2}\left(\frac{d\phi}{dt}\right)^{2}  = - \frac{\frac{R}{2}\left(1 +
\tilde{w}\right)}{1 + \frac{R}{2}\left(1 +
\tilde{w}\right)}\frac{G}{4l^{2}}\,\sigma^{4}F
 \ .\label{kinF}
\end{equation}
Consequently, the energy density (\ref{rho1}) becomes
\begin{equation}
\rho_{\phi} =  \frac{1}{1 + \frac{R}{2}\left(1 +
\tilde{w}\right)}\frac{G}{4l^{2}}\,\sigma^{4}F = - \frac{2}{R\left(1 +
\tilde{w}\right)}\sigma^{2}\left(\frac{d\phi}{dt}\right)^{2}
 \ .\label{rphi}
\end{equation}
The expression (\ref{rphi}), hence the total energy density
(\ref{rhoeffdec}) and the energy density (\ref{rho2}), are
positive for any positive $F$ and $\tilde{w}> -1 $ as long as
$\sigma^{2}<0$. Consequently, although $\sigma^{2}<0$, equivalent
to a negative kinetic term, the energy density of each of the
components as well as the total energy density are always
positive.
For $w_{\phi} = 0$, i.e. $\tilde{w} < -1$, we have $\sigma^{2}>0$ and
\begin{equation}
\rho_{\phi} = 2\sigma^{2}\left(\frac{d\phi}{dt}\right)^{2} \qquad\qquad (w_{\phi}=0)
 \ .\label{rphi+}
\end{equation}
The pressure (\ref{p1}) becomes
\begin{equation}
p_{\phi} =  - \frac{1 + R\left(1 + \tilde{w}\right)}{1 +
\frac{R}{2}\left(1 + \tilde{w}\right)}\frac{G}{4l^{2}}\,\sigma^{4}F =
2 \frac{1 + R\left(1 + \tilde{w}\right)}{R\left(1 +
\tilde{w}\right)}\sigma^{2}\left(\frac{d\phi}{dt}\right)^{2}
 \ ,\label{pphi}
\end{equation}
together with (\ref{rphi}) consistently reproducing the EoS parameter (\ref{rhoconst}). Thus, the Einstein frame dynamics of the system is completely solved for the given configuration. In the following subsection we shall use the transformation relations (\ref{conv}) and (\ref{tH}) to obtain the corresponding physical dynamics.

\subsection{Dynamics in the physical frame}

The transformations from Einstein`s to the physical frame are
mediated by exponentials of $\phi$.  With (\ref{ephi})
 one has $\phi = $
const for $\tilde{w} = - 1$.
For the physical scale factor, relations (\ref{conv}) and
(\ref{ephi}) provide us with
\begin{equation}
\tilde{a} = e^{-\phi}a \quad \Rightarrow \quad \tilde{a} =
a^{-\frac{2}{1 + 3 \tilde{w}}}
 \  .\label{tila}
\end{equation}
For any $\tilde{w} < - \frac{1}{3}$, the scale factor $\tilde{a}$ is proportional to a positive power of $a$.
It is only for $\tilde{w} = -1$ that $\tilde{a} = a$. For
non-relativistic matter we find
\begin{equation}
\tilde{a} = \frac{1}{a^{2}} \quad \Rightarrow\quad a =
\frac{1}{\tilde{a}^{1/2}} \qquad\qquad (\tilde{w}=0)
 \  .\label{tilaw0}
\end{equation}
In this case, a growing (decaying) $a$ generates a decaying
(growing) $\tilde{a}$. In particular, to have an increasing scale
factor $\tilde{a}$, the corresponding $a$ is the scale factor of a
contracting solution.
For $w_{\phi} = 0$, equivalent to $\tilde{w} < -1$, we have
\begin{equation}
\tilde{a} = a^{\frac{1}{1+3/(2R)}}\quad \Rightarrow\quad a = \tilde{a}^{\left(1+3/(2R)\right)}
\qquad\qquad (w_{\phi}=0)
 \  .\label{tilaw0+}
\end{equation}
A growing solution always remains a growing solution.
Generally, we have $\tilde{\rho} = \rho_{f}\exp\left[2\phi\right]$. With $\rho_{f} = $ const (cf. (\ref{rhopropto1}) with (\ref{rhoconst})), it follows that
\begin{equation}
\tilde{\rho} \propto \tilde{a}^{-3\left(1 + \tilde{w}\right)}
 \  .\label{tilrho}
\end{equation}
This dependence is consistent
with the balance (\ref{tildecons}). Notice that
this is the dynamics of the matter component under the condition
of a total EoS $\frac{p_{eff}}{\rho_{eff}} = - 1$,
where the matter energy density  $\rho_{m}$ via (\ref{rho2}) and
(\ref{rhoeffdec}) is part of the total energy density
$\rho_{eff}$. It is not just the dynamics on a given background.

In a next step we consider the relation (\ref{tH}) between
$\tilde{H}$ and $H$. With (\ref{dotphiHfin}) and (\ref{ephi})
it follows that
\begin{equation}
H - \frac{d\phi}{dt} = - \frac{2}{1 + 3 \tilde{w}}H \quad \Rightarrow\quad
\tilde{H} =  - \frac{2}{1 + 3 \tilde{w}}\, a^{-
3\frac{1+\tilde{w}}{1+3\tilde{w}}}\,H
 \  .\label{HtH-}
\end{equation}
Recall that $\tilde{w}\neq-1/3$.
For any $\tilde{w}>-1/3$ the rates $\tilde{H}$ and $H$ have different
signs. Then, a contracting $H$ corresponds to an expanding $\tilde{H}$
and vice versa. The EoS $\tilde{w}=-1$ is the only case
for which $\tilde{H}=H$. For non-relativistic matter the
connection between both Hubble rates is
\begin{equation}
\tilde{H} =  - \frac{2H}{a^{3}} = - 2 \tilde{a}^{3/2}\,H\qquad\qquad (\tilde{w}=0)
 \  .\label{HtH-0}
\end{equation}
For $H= $ const $<0$, the physical Hubble rate increases with
$\tilde{a}^{3/2}$. This demonstrates again the phantom character
of the solution.
On the other hand, for $w_{\phi}=0$, equivalent to a phantom-type EoS of the matter, we obtain
\begin{equation}
\tilde{H} =  \frac{1}{1+3/(2R)}a^{-1/(1+2R/3)}
\,H\qquad\qquad (w_{\phi}=0)
 \  .\label{HtH-0+}
\end{equation}
In this case, $H$ and $\tilde{H}$ have always the same sign.
An explicit expression of $\tilde{a}$ in terms of
$\tilde{t}$ is obtained with the help of (\ref{HtH-}) with
(\ref{tila}). Then,
\begin{equation}
\tilde{H} = - \frac{2}{1 + 3 \tilde{w}}\tilde{a}^{\frac{3}{2}\left(1 +
\tilde{w}\right)}H \quad \Rightarrow\quad \frac{d \tilde{a}}{d
\tilde{t}} = - \frac{2}{1 + 3 \tilde{w}}\tilde{a}^{\frac{1}{2}\left(5
+ 3\tilde{w}\right)}H
 \   \label{tHta}
\end{equation}
with a constant $H$. The integration yields
\begin{equation}
\tilde{a} = \frac{\tilde{a}_{i}}{\left[1 + 3\frac{1 + \tilde{w}}{1 + 3
\tilde{w}}\tilde{a}_{i}^{\frac{3}{2}\left(1 +
\tilde{w}\right)}\,H\,\left(\tilde{t} -
\tilde{t}_{i}\right)\right]^{\frac{2}{3\left(1 + \tilde{w}\right)}}}
 \  . \label{taf}
\end{equation}
We mention that a solution $\tilde{a} \propto \tilde{t}^{-\frac{2}{3\left(1 + \tilde{w}\right)}}$ has been obtained before in \cite{ratra}.
According to (\ref{tHta}), an expanding solution for the physical Hubble rate, $\tilde{H}>0$, corresponds to $H<0$ for $\tilde{w}>-\frac{1}{3}$. But with $H<0$ and $\tilde{w}>-\frac{1}{3}$ the denominator in (\ref{taf}) vanishes for $\tilde{t} > \tilde{t}_{i}$.
The solution (\ref{taf})
describes an expanding universe that approaches a singularity
after a finite time (big rip). The time span is determined by the
vanishing of the denominator in (\ref{taf}). For the case of
non-relativistic matter (\ref{taf}) reduces to
\begin{equation}
\tilde{a} = \frac{\tilde{a}_{i}}{\left[1 +
3\tilde{a}_{i}^{\frac{3}{2}}\,H\,\left(\tilde{t} -
\tilde{t}_{i}\right)\right]^{\frac{2}{3}}}\qquad\qquad (\tilde{w}= 0)
 \  . \label{taf0}
\end{equation}
The big rip singularity is approached after a time span
$\tilde{t}_{br}$
\begin{equation}
1 + 3\tilde{a}_{i}^{\frac{3}{2}}\,H\,\left(\tilde{t}_{br} -
\tilde{t}_{i}\right) = 0\quad \Rightarrow\quad \tilde{t}_{br} =
\tilde{t}_{i} + \frac{1}{3 \tilde{a}_{i}^{\frac{3}{2}}|H|}
 \  . \label{br}
\end{equation}
Notice that this big rip solution is different from Caldwell's solution in GR \cite{caldwell}. There, it is a consequence of $\tilde{w} < -1$. In our case it follows for $\tilde{w}>-\frac{1}{3}$ which implies a contracting solution $H<0$ in the Einstein frame. No singularity occurs for $\tilde{w} < -1$.
For $w_{\phi}=0$, i.e., $\tilde{w} < -1$ and $H> 0$,  we find
\begin{equation}
\tilde{a} = \tilde{a}_{i}\left[1 + \frac{1}{1+2R/3}\tilde{a}_{i}^{-3/(2R)}H\left(\tilde{t}-\tilde{t}_{i}\right)\right]^{2R/3}
\qquad\qquad (w_{\phi}= 0)
 \  . \label{taf0+}
\end{equation}
Accelerated expansion requires $R>3/2$.

The time coordinates $t$ and $\tilde{t}$ are related by the first transformation of (\ref{conv}).
With (\ref{ephi}) we find
\begin{equation}
\tilde{t} \propto e^{3H\frac{1+\tilde{w}}{1+3\tilde{w}}t} \quad \Rightarrow\quad Ht \propto \ln \tilde{t}^{\frac{1}{3}\frac{1+3\tilde{w}}{1+\tilde{w}}}
 \   \label{ttilt}
\end{equation}
as well as
\begin{equation}
e^{-\phi} \propto \frac{1}{\tilde{t}} \quad \Rightarrow\quad \tilde{a} \propto \frac{a}{\tilde{t}}
 \ .  \label{ephitil}
\end{equation}
Combining $a \propto \exp\left[Ht\right]$ with (\ref{ttilt}) and (\ref{ephitil}), it follows consistently that
\begin{equation}
\tilde{a} \propto \tilde{t}^{\frac{2}{3\left(1+\tilde{w}_{eff}\right)}}
 \ ,   \label{tatt}
\end{equation}
where we have introduced the physical EoS parameter
\begin{equation}
\tilde{w}_{eff} = - 2 - \tilde{w}
 \ . \label{tweff-}
\end{equation}
For any $\tilde{w} > -1$, including the case $\tilde{w} =0$, the total equation of state is of the phantom type.
For $\tilde{w} = -1$ it follows $\tilde{w}_{eff} = -1$.
It is only for $\tilde{w} = -2$ that we recover a total matter-type behavior $\tilde{w}_{eff} = 0$, which, according to (\ref{rhoconst}), implies $w_{\phi} = - 1 +R$.
A radiation behavior requires $\tilde{w} = -\frac{7}{3}$.
For $w_{\phi}= 0$, i.e., $\tilde{w} < -1$,  the parameter $\tilde{w}_{eff}$ becomes
\begin{equation}
\tilde{w}_{eff} = - 1 + \frac{1}{R}\qquad\qquad (w_{\phi} = 0)
 \ . \label{tweffR}
\end{equation}
The medium as a whole can be seen as characterized by a physical energy density
\begin{equation}
\tilde{\rho}_{eff}  \propto \tilde{a}^{-3\left(1 + \tilde{w}_{eff}\right)}
 \  ,\label{treff}
\end{equation}
as solution of the conservation equation
\begin{equation}
\frac{d\tilde{\rho}_{eff}}{d\tilde{t}}  + 3\tilde{H}\left(1 + \tilde{w}_{eff}\right)\tilde{\rho}_{eff} = 0
 \  .\label{constil}
\end{equation}
With (\ref{tweff-}), the dynamics of the physical frame is explicitly known. The cosmic substratum as a whole behaves like a fluid with an effective EoS $\tilde{w}_{eff}$.
The relation between the physical EoS parameter $\tilde{w}_{eff}$ and the auxiliary quantity $w_{eff} = -1$ is different from a corresponding relation in scalar-tensor theories \cite{EdWi}. In scalar tensor theories, both frames are related by a conformal transformation which is not the case in TeVeS.

The deceleration parameter in the physical frame is
\begin{equation}
\tilde{q} \equiv - \frac{\tilde{a}\,\frac{d^{2} \tilde{a}}{d
\tilde{t}^{2}}}{\left(\frac{d \tilde{a}}{d \tilde{t}}\right)^{2}}
=  - \frac{\left(1 - \frac{2}{H} \frac{d\phi}{dt}\right) \left(1 -
\frac{1}{H} \frac{d\phi}{dt}\right) + \frac{1}{H^{2}}\frac{dH}{dt} -
\frac{1}{H^{2}}\frac{d^{2}\phi}{dt^{2}}}{\left(1 -
\frac{1}{H}\frac{d\phi}{dt}\right)^{2}}
 \ . \label{qt1}
\end{equation}
It is related to the corresponding parameter $
q = - \frac{1}{a H^{2}}\frac{d^{2}a}{dt^{2}}$
in the Einstein frame by
\begin{equation}
\tilde{q} = q  - \frac{\frac{1}{H}\frac{d\phi}{dt}\left(1 -
\frac{1}{H}\frac{d\phi}{dt}\right)\left(\frac{1}{H^{2}}\frac{dH}{dt} - 1\right)
-
\frac{1}{H}\frac{d}{dt}\left(\frac{1}{H}\frac{d\phi}{dt}\right)}
{\left(1
- \frac{1}{H}\frac{d\phi}{dt}\right)^{2}}
 \ . \label{qqE}
\end{equation}
We have accelerated expansion if the numerator of the right-hand
side of eq.~(\ref{qt1}) is positive,
\begin{equation}
\tilde{q} < 0 \quad\Leftrightarrow \quad  \left(1 - \frac{2}{H} \frac{d\phi}{dt}\right) \left(1 -
\frac{1}{H} \frac{d\phi}{dt}\right) + \frac{1}{H^{2}}\frac{dH}{dt} -
\frac{1}{H^{2}}\frac{d^{2}\phi}{dt^{2}} > 0
 \ . \label{condacc}
\end{equation}
For the dynamics discussed in this paper ($\tilde{w} = $ const,
$w_{\phi} = $ const, $R = $ const) we have $q = - 1$. With
$\frac{dH}{dt} = \frac{d^{2}\phi}{dt^{2}} = 0$ in eq.~(\ref{qt1}), use of
eq.~(\ref{dotphiHfin}) provides us with
\begin{equation}
\tilde{q} = - \frac{5 + 3\tilde{w}}{2} = \frac{1}{2}\left(1 + 3 \tilde{w}_{eff}\right)
 \ . \label{qfin}
\end{equation}
Consistently, we have $\tilde{q} = -1$ for $\tilde{w} = -1$. For
non-relativistic matter $
\tilde{q} = - \frac{5}{2}$
is valid. In this case $\tilde{q}$
does not depend on $R$. For $w_{\phi} = 0$, i.e. $\tilde{w} < -1$,  the deceleration parameter becomes
\begin{equation}
\tilde{q} = - \left(1 - \frac{3}{2R}\right) \qquad\qquad\qquad (w_{\phi} = 0)\ .
 \  \label{qfin0+}
\end{equation}
Accelerated expansion is obtained for $R > \frac{3}{2}$ in accordance with (\ref{taf0+}). The deceleration parameter vanishes for $R = \frac{3}{2}$ which corresponds to $\tilde{w}_{eff} = - \frac{1}{3}$ (cf. Eq.~(\ref{tweffR})). For $R = 1$, equivalent to $\tilde{w} = -2$, we consistently recover $\tilde{q} = \frac{1}{2}$.

\subsection{Implications for the $F$ function}

Now let us combine  Eqs.~(\ref{consist}) and (\ref{dotphiHfin}). The
result is
\begin{equation}
\frac{G}{4l^{2}}\,\frac{d}{ \sigma^{2}}\left(\sigma^{4}F\right)
 = 9\,\frac{\left(1 +
\tilde{w}\right)^{2}}{\left(1 + 3\tilde{w}\right)^{2}} H^{2}
 \ . \label{mu2Fp}
\end{equation}
Then, the relations (\ref{mu2Fp}) and  (\ref{FriedmR1}) provide us with
\begin{equation}
\frac{\frac{d}{d \sigma^{2}}\left(\sigma^{4}F\right)}{\sigma^{4}F}
= 24\pi\,G \frac{\left(1 + R\right)\left(1 +
\tilde{w}\right)^{2}}{\left(1 + 3\tilde{w}\right)^{2}\left[1 +
\frac{R}{2}\left(1 + \tilde{w}\right)\right]}
 \ . \label{mu2Fpmu2}
\end{equation}
Using the expression (\ref{mu/kfin}) in (\ref{mu2Fpmu2}), the latter takes the
form
\begin{equation}
\frac{\frac{d}{d \sigma^{2}}\left(\sigma^{4}F\right)}{\sigma^{4}F}
= - \frac{R}{2\sigma^{2}}\frac{1 + \tilde{w}}{1 + \frac{R}{2}\left(1 +
\tilde{w}\right)}
 \  \label{mu2Fpmu2mu}
\end{equation}
with the solution
\begin{equation}
F \propto \left(\sigma^{2}\right) ^{-\frac{4 + 3\left(1+\tilde{w}\right)R}{2 + \left(1+\tilde{w}\right)R}}
 \ . \label{Fsol}
\end{equation}
For the special case $\tilde{w}=0$  with $\sigma^{2} < 0$  we have
\begin{equation}
F \propto \left(-\sigma^{2}\right) ^{-\frac{3R+4}{R+2}}
\qquad\qquad (\tilde{w}=0)
 \ . \label{Fpropto}
\end{equation}
For $w_{\phi}=0$, equivalent to $\tilde{w} < -1$, the dependence of $F$ is
\begin{equation}
F \propto \frac{1}{\sigma^{2}}
\qquad\qquad (w_{\phi}=0)
 \ . \label{Fpropto+}
\end{equation}
This is consistent with $w_{\phi}=0$ in (\ref{w1}) under the condition $\frac{d\phi}{dt} =$ const.

While in general there does not exist a theory for Bekenstein's
$F$ function, we found that for the special situation considered
in this paper, there is no remaining freedom for the choice of
this function. The existence of a scaling solution requires a
structure that is different from that suggested by Bekenstein.
It remains open, whether the dependence
(\ref{Fsol}) is a large scale  limit of a more general
function.
We mention that in the vacuum limit $R=0$ the relation
(\ref{Fpropto}) implies $\sigma^{4}F = $ constant which proves the
statement following eq.~(\ref{ddphir2+}). This is, together with
$\sigma^{2} = 0$ for $R=0$, consistent with an EoS
$w_{\phi}= - 1$ (cf. eqs.~(\ref{w1}) and (\ref{rhoconst})).
It is also interesting to realize that the
structure of (\ref{Fpropto}) is similar to that of an additional
(to Bekenstein's ansatz for $F$) term introduced ad hoc in
\cite{hao}. For the special case $R=0$ the resulting inverse
quadratic dependence on $-\sigma^{2}$ in (\ref{Fpropto}) even
exactly (up to the sign) coincides with the structure of that term
(cf. eq.~(21) in \cite{hao}).

\section{The cosmic medium}
\label{cosmed}

The energy density (\ref{treff}) can be seen as the solution of a conservation equation (\ref{constil})
with an effective pressure
\begin{equation}
\tilde{p}_{\mathrm{eff}} = \tilde{w}_{\mathrm{eff}}\tilde{\rho}_{\mathrm{eff}}
 \ ,\label{peff}
\end{equation}
where $\tilde{w}_{\mathrm{eff}}$ is given by (\ref{tweff-}). Then
\begin{equation}
\tilde{\rho}_{\mathrm{eff}} \propto  \tilde{a}^{3\left(1 + \tilde{w}\right)}
 \ .\label{trtw}
\end{equation}
The overall dynamics is completely determined by the EoS parameter $\tilde{w}$ of the matter. Does this solution have any relevance for our real Universe? As already mentioned, for any $\tilde{w} > - 1$ the effective energy density $\tilde{\rho}_{\mathrm{eff}}$ increases with $\tilde{a}$, i.e., the total dynamics is of the phantom type. If, on the other hand, $\tilde{w}$ itself is of the phantom type, i.e., $\tilde{w} < - 1$, one has $\tilde{w}_{\mathrm{eff}} > -1$ for the medium as a whole. The matter dynamics is described by (cf. Eq.~(\ref{tildecons}))
\begin{equation}
\tilde{\rho} \propto  \tilde{a}^{-3\left(1 + \tilde{w}\right)}\
 \ .\label{tmat}
\end{equation}
Equation (\ref{constil}) with the solution (\ref{treff}) and (\ref{tweff-}) represents a GR equivalent for  the TeVeS dynamics.
Given the separately conserved matter component with energy density $\tilde{\rho}$ and the conserved total energy with a density $\tilde{\rho}_{\mathrm{eff}}$, one may associate a so far unknown component with an energy density $\tilde{\rho}_{y}$ to the difference between $\tilde{\rho}_{\mathrm{eff}}$ and $\tilde{\rho}$,
\begin{equation}
\tilde{\rho}_{\mathrm{eff}} = \tilde{\rho} + \tilde{\rho}_{y}
 \ .\label{sr}
\end{equation}
Let us further assume that the geometry based $y$-component can be described by an effective (not necessarily constant) EoS parameter $\tilde{w}_{y}$ such that
\begin{equation}
\frac{d\tilde{\rho}_{y}}{d\tilde{t}} + \left(1 + \tilde{w}_{y}\right)\tilde{\rho}_{y} = 0
 \ .\label{consx}
\end{equation}
In an effective GR description, the Universe is now made up of two separately conserved components with energy densities $\tilde{\rho}$ and $\tilde{\rho}_{y}$.
Then
\begin{equation}
\tilde{w}_{\mathrm{eff}} = \tilde{w}\frac{\tilde{\rho}}{\tilde{\rho}_{\mathrm{eff}}} + \tilde{w}_{y}\frac{\tilde{\rho}_{y}}{\tilde{\rho}_{\mathrm{eff}}}
 \ .\label{decw}
\end{equation}
Introducing the ratio
\begin{equation}
\tilde{r} \equiv \frac{\tilde{\rho}}{\tilde{\rho}_y}
 \ ,\label{tr}
\end{equation}
we may write
\begin{equation}
\tilde{w}_{y} = \left(1 + \tilde{r}\right) \tilde{w}_{\mathrm{eff}} - \tilde{r}\tilde{w}
 \ .\label{twx}
\end{equation}
Since
\begin{equation}
\tilde{r} = \frac{\tilde{\rho}}
{\tilde{\rho}_{\mathrm{eff}} - \tilde{\rho}
}
= \frac{\frac{\tilde{\rho}}{\tilde{\rho}_{\mathrm{eff}}}}{1 - \frac{\tilde{\rho}}{\tilde{\rho}_{\mathrm{eff}}}} \qquad \mathrm{and} \qquad
1 + \tilde{r} = \frac{1}{1 - \frac{\tilde{\rho}}{\tilde{\rho}_{\mathrm{eff}}}}
 \ ,\label{tr1}
\end{equation}
and, since
the ratio $\frac{\tilde{\rho}}{\tilde{\rho}_{\mathrm{eff}}}$ is known,
\begin{equation}
\frac{\tilde{\rho}}{\tilde{\rho}_{\mathrm{eff}}} =  \tilde{\Omega}_{0}\tilde{a}^{-6\left(1 + \tilde{w}\right)} \qquad \mathrm{with} \qquad\tilde{\Omega}_{0} \equiv \frac{\tilde{\rho}_0}{\tilde{\rho}_{\mathrm{eff}0}}
 \ ,\label{ratiotr}
\end{equation}
we obtain
\begin{equation}
\tilde{r} = \frac{\tilde{\Omega}_{0}
\tilde{a}^{-6\left(1 + \tilde{w}\right)}}
{1 - \tilde{\Omega}_{0}\tilde{a}^{-6\left(1 + \tilde{w}\right)}} \qquad \mathrm{and} \qquad
1 + \tilde{r} = \frac{1}{1 - \tilde{\Omega}_{0}\tilde{a}^{-6\left(1 + \tilde{w}\right)}}
 \ .\label{trta}
\end{equation}
For $\tilde{w}_{y}$ we find
\begin{equation}
\tilde{w}_{y} = \tilde{w}_{\mathrm{eff}} -2\tilde{r} \left(1 + \tilde{w}\right) = - 2 - \tilde{w} -2\tilde{r} \left(1 + \tilde{w}\right)
 \ .\label{twx1}
\end{equation}
In terms of the scale-factor,
\begin{equation}
\tilde{w}_{y} = - \frac{
2 +\tilde{w}\left(1 + \tilde{\Omega}_{0}\tilde{a}^{-6\left(1 + \tilde{w}\right)}\right)}
{1 - \tilde{\Omega}_{0}\tilde{a}^{-6\left(1 + \tilde{w}\right)
}}
 \ .\label{twx}
\end{equation}
As already mentioned, equations of state $\tilde{w}_{eff} = \frac{1}{3}$ and $\tilde{w}_{eff} = 0$ can be reproduced by $\tilde{\omega}=-\frac{7}{3}$ and $\tilde{\omega}=-2$, respectively.
For $\tilde{w}_{eff} = \frac{1}{3}$ relation (\ref{twx}) specifies to
\begin{equation}
\tilde{w}_{y} =  \frac{1}{3}\frac{ 1 + 7\tilde{\Omega}_{0}\tilde{a}^{8}}
{1 - \tilde{\Omega}_{0}\tilde{a}^{8}} \qquad \qquad (\tilde{w}_{eff} = \frac{1}{3})
 \ .\label{twy/3}
\end{equation}
For $\tilde{a} \ll 1$, i.e., at high redshifts, the EoS parameter $\tilde{w}_{y}$ approaches $\tilde{w}_{y}=\frac{1}{3}$. The $y$ component, an entity of geometric origin, behaves as radiation as well.
With respect to the DM problem, the case $\tilde{w}_{eff} = 0$ is of particular interest.
For this case Eq.~(\ref{twx}) takes the form
\begin{equation}
\tilde{w}_{y} =  \frac{
2\tilde{\Omega}_{0}\tilde{a}^{6}}
{1 - \tilde{\Omega}_{0}\tilde{a}^{6}} \qquad \qquad (\tilde{w}_{eff} = 0)
 \ .\label{twy}
\end{equation}
For $\tilde{a} \ll 1$ the EoS parameter $\tilde{w}_{y}$ approaches $\tilde{w}_{y}=0$. The $y$ component behaves as non-relativistic matter. So, the apparently strange result is, that to have a non-relativistic matter period  $\tilde{w}_{eff} = 0$ we need a $y$ component with $\tilde{w}_{y}=0$ that only exists in the presence of a phantom type DE component with $\tilde{w}=-2$. The total energy density is the sum of the energy densities of these components. With $\tilde{w}=-2$ the EoS of the scalar field component is $w_{\phi} = -1 + R$. It is only a combination of $w_{\phi}$ and $R$ which is fixed. The apparently most natural choice $w_{\phi} = 0$ is realized for $R=1$. But for the physical ratio $\tilde{r}$ we obtain
\begin{equation}
\tilde{r} = \frac{\tilde{\Omega}_{0}
\tilde{a}^{6}}
{1 - \tilde{\Omega}_{0}\tilde{a}^{6}}  \qquad \qquad (\tilde{w}_{eff} = 0)
 \ ,\label{trta+}
\end{equation}
from (\ref{trta}), i.e., $\tilde{r} \rightarrow 0$ for $\tilde{a} \ll 1$. This is equivalent to
\begin{equation}
\tilde{\rho}_{y} \gg \tilde{\rho}\qquad\Rightarrow \qquad \tilde{\rho}_{eff} \approx \tilde{\rho}_{y}
\qquad \qquad (\tilde{a} \ll 1)
 \ ,\label{ry>}
\end{equation}
as one expects for a matter-dominated universe. The DE component $\tilde{\rho}$ is dynamically negligible at high redshifts, but its presence is crucial for the non-relativistic matter component to exist.
These results demonstrate, that physical equations of state such as $\tilde{w}_{eff} = \frac{1}{3}$ or $\tilde{w}_{eff} = 0$ can be obtained within TeVeS on a purely geometric basis, provided only that there exists a separate phantom-type DE component.
Finally we also mention that for $\tilde{w}=-1$ which implies $\tilde{w}_{eff} = 1$ we obtain $\tilde{w}_{y}=-1$ as well.

At the present epoch (\ref{twx}) reduces to
\begin{equation}
\tilde{w}_{y0} = - \frac{
2 +\tilde{w}\left(1 + \tilde{\Omega}_{0}\right)
}
{1 - \tilde{\Omega}_{0}
}
 \ .\label{twx0}
\end{equation}
Since we have shown, that there exists a solution with $\tilde{w}_{y}=0$ at high redshifts, it seems tempting to consider the possibility of a present value $\tilde{w}_{y0} \approx 0$ as well.
This corresponds to the original intention of TeVeS to avoid the introduction of a DM component.
Under the condition $\tilde{w}_{y0} \approx 0$ we have from Eq.~(\ref{twx0}),
\begin{equation}
\tilde{w}  \approx - \frac{2}
{1 + \tilde{\Omega}_{0}}
 \ .\label{twx0=0}
\end{equation}
If the $y$-component is to play the role of DM today, the ``matter" component with an EoS $\tilde{w}$ has again to be associated with DE in an equivalent GR picture. As already mentioned, this is in accordance with the findings in \cite{ratra} where the authors obtained an inflationary solution
for $\tilde{w} < - 1$. Now we know that the actual background universe is quite well described by the $\Lambda$CDM model. Within the latter, the total EoS parameter is
\begin{equation}
\tilde{w}_{eff}^{\Lambda \mathrm{CDM}} = - \frac{1}{1+\rho_{M0}/\rho_{\Lambda}}
 \ ,\label{eosla}
\end{equation}
where $\frac{\rho_{M0}}{\rho_{\Lambda}} \approx \frac{1}{3}$. This corresponds to a value
$\tilde{\Omega}_{0} \approx \frac{3}{4}$. Using this ratio in (\ref{twx0=0}), we find an effective EoS for the DE equivalent of $\tilde{w} \approx - \frac{8}{7} = - 1.14$. Such a value is compatible with current observational data \cite{vikhlinin,sanchez}.
Under the condition (\ref{twx0=0}) it follows from (\ref{rhoconst}) that
\begin{equation}
w_{\phi} \approx -1 + R\frac{1 - \tilde{\Omega}_{0}}{1 + \tilde{\Omega}_{0}} \approx -1 + \frac{R}{7}
 \ .\label{}
\end{equation}
It is only a combination of $w_{\phi}$ and $R$ which is fixed, not each of these quantities by itself.  However, the value for $\tilde{w}$ is consistent with $w_{\phi} =0$ which requires $R=7$.
With this choice it is obvious that the scalar field represents the DM.

These considerations show that different epochs of the cosmological evolution are reproduced for different values of $\tilde{w}$. An early radiation period requires $\tilde{w} = -\frac{7}{3}$, a subsequent matter period has $\tilde{w} = -2$ and for the present epoch we found $\tilde{w} \approx -\frac{8}{7}$. This indicates a tendency to less negative values of $\tilde{w}$ with the cosmic time. Guided by the $\Lambda$CDM model, one may speculate that in the future $\tilde{w}$ approaches $\tilde{w} = -1$, which means $\tilde{w}_{eff} = -1$ as well. As in the $\Lambda$CDM model, the final state of the universe would be of the de Sitter type. As already mentioned, the parameter $\tilde{w}_{y}$, according to (\ref{twx1}), also tends to $\tilde{w}_{y}=-1$.
In this limit, all the components, including the $y-$ component, behave as DE. The final state of the Universe would be characterized by two different DE contributions, one from the matter sector, the other one of geometric origin. However, since power-law solutions do not allow to construct a scenario, this remains speculative. Perhaps a dynamical system analysis could clarify whether there exist fixed points with corresponding properties.
On the other hand, any value of $\tilde{w}$ different from $\tilde{w} = -1$ in the long-time limit would invalidate the split (\ref{sr}). Namely, for any $\tilde{w} < -1$ the ratio $\frac{\tilde{\rho}}{\tilde{\rho}_{eff}}$ in (\ref{ratiotr}) would grow to values larger than unity, which implies a negative energy density $\tilde{\rho}_{y}$ together with a negative value of $\tilde{r}$ in (\ref{trta}). Consequently, a limit $\tilde{w} = -1$ is simply necessary for our power-law solution to be applicable for the future universe with $a>1$.

\section{Discussion}
\label{discussion}

We have solved a simple case of the TeVeS dynamics in a
homogeneous, isotropic and spatially flat universe. The total
effective energy density was split into two components in a  way
that allowed us to look for scaling solutions, i.e., solutions
with a constant ratio of the energy densities of these components.
One of the components is a scalar field with a potential part, related to Bekenstein's $F$ function,
the other one represents a ``fluid" contribution.
TeVeS dynamics in the ``Einstein frame" then corresponds to an interacting two-component dynamics with explicitly given coupling term.
The Einstein frame Hubble parameter is necessarily constant and Bekenstein's $F$ function is fixed. For the physical Hubble parameter we find
a power-law solution, corresponding to a total effective energy density $\tilde{\rho}_{eff}$ with an EoS parameter $\tilde{w}_{eff} = - 2 - \tilde{w}$, where $\tilde{w}$ is the ``matter" EoS parameter. ``Matter" is understood here in a general sense which includes $\tilde{w} < 0$ as well.
We establish an effective GR picture of the dynamics according to which the cosmic substratum consists of two separately conserved components: the first one is the mentioned ``matter" with energy density $\tilde{\rho}$ and EoS parameter $\tilde{w}$, the second one is characterized by an energy density $\tilde{\rho}_{y} \equiv \tilde{\rho}_{eff} - \tilde{\rho}$ of geometrical origin and a (generally not constant) EoS parameter $\tilde{w}_{y}$ which is calculated from the explicitly known dynamics of $\tilde{\rho}_{eff}$ and $\tilde{\rho}$.
A non-relativistic matter period of the cosmic expansion can only be obtained for $\tilde{w} = -2$. The ``matter" has to represent (phantom type) dark energy.
For small values of the scale factor we find a non-relativistic matter-type phase $\tilde{w}_{eff} = 0$ with $\tilde{w}_{y} \rightarrow 0$, i.e.,
the geometry based component acts as matter. Moreover, the dark energy fluid is dynamically irrelevant in this limit. However, its mere existence is absolutely necessary to obtain a realistic dynamics. This demonstrates that an equation of state for  non-relativistic matter can be obtained within TeVeS on a purely geometrical basis.
For the actual Universe the situation is less clear. But, of course, it has to be approximated by a different power-law solution. Assuming tentatively the $\Lambda$CDM motivated
present value $\frac{\tilde{\rho}_{0}}{\tilde{\rho}_{eff,0}} \approx \frac{3}{4}$ together with $\tilde{w}_{y0} = 0$, this requires a present dark energy EoS parameter of $\tilde{w} \approx -1.14$, which seems not yet ruled out by the data.
In a sense, this picture, although perhaps surprising, is consistent with the original intention of MOND type theories, to avoid the introduction of a dark matter component. While our solution implies a geometrization of dark matter, the existence of a separate dark energy component is not only necessary to obtain the observed amount of accelerated expansion but also to recover a  non-relativistic matter phase of the cosmic medium.  Future investigations along these lines will have to include baryon and radiation components as well as perturbations about the homogeneous and isotropic background.

\vspace{1.0cm}
%%%%%%%%%%%%%%%%%%%%%%%%%%%%%%%%%%%%%%%%%%%%%%%%%%%%%%%%%%%%
\noindent
{\bf Acknowledgement:} Financial support by FAPES and CNPq (Brazil) is gratefully acknowledged.
%%%%%%%%%%%%%%%%%%%%%%%%%%%%%%%%%%%%%%%%%%%%%%%%%%%%%%%%%%%%

\end{document}